# Strong Coupling Lattice Schwinger Model on Large Spherelike Lattices

**H. Gausterer and C.B. Lang**

Institut für Theoretische Physik
Universität Graz
A-8010 Graz, AUSTRIA

**Abstract**

The lattice regularized Schwinger model for one fermion flavor and in the strong coupling limit is studied through its equivalent representation as a restricted 8-vertex model. The Monte Carlo simulation on lattices with torus-topology is handicapped by a severe non-ergodicity of the updating algorithm; introducing lattices with spherelike topology avoids this problem. We present a large scale study leading to the identification of a critical point with critical exponent $\nu = 1$, in the universality class of the Ising model or, equivalently, the lattice model of free fermions.

# 1 Motivation

Several models of lattice regularized quantum field theories with fermions may be expressed by simple statistical systems of monomer-dimer type. Some time ago it was proven [1] that the lattice regularized version of $\mathrm{QED}_2$, i.e. compact $U(1)$ gauge theory with one flavor of fermions (the Schwinger model [2]) in the Wilson formulation

$$\begin{aligned} S &= S_F[\bar{\psi}, \psi, U] + \beta S_G[U] \;, \\ S_F &= \sum_{x\in\Lambda} \left( \frac{1}{2} \sum_{\mu} \Big( \bar{\psi}(x+\hat{\mu})(1+\gamma_\mu) U_\mu^\dagger(x) \psi(x) \right. \\ &\quad \left. + \bar{\psi}(x)(1-\gamma_\mu) U_\mu(x) \psi(x+\hat{\mu}) \Big) - \bar{\psi}(x) M \psi(x) \right) \;, \end{aligned} \tag{1}$$

– where $U$ denotes the gauge link variables and $S_G$ may be any gauge field action (compact or non-compact) – is in the limit $\beta \to 0$ equivalent to a specific 8-vertex model. Such a formulation is useful since it allows to explore the phase structure with alternative methods. In particular, the crucial question in this model is whether there is a phase transition at non-zero but finite values of the fermion mass parameter and if yes, what type it may be.

To be more explicit, in [1] it was shown that at strong coupling ($\beta = 0$) the partition function of the above model agrees with the partition function of the 8-vertex model

$$Z_\Lambda = Z_{8V} = \sum_{\{n_i\}} \prod_{i=1}^{8} a_i^{n_i} \quad , \tag{2}$$

where $n_i$ and $a_i$ denote the multiplicity and corresponding weight of the vertices of type $i$. For general details on the 8-vertex model see [3]. The specific weights in (2) are

$$a_1 = M^2 \equiv \frac{1}{(2\kappa)^2}, \; a_2 = 0, \; a_3 = a_4 = 1, \; a_i = \frac{1}{2} \text{ (for } i \geq 5). \tag{3}$$

(We have introduced here the frequently used "hopping" parameter $\kappa$.) A certain definition of the vertices in agreement with [3] is

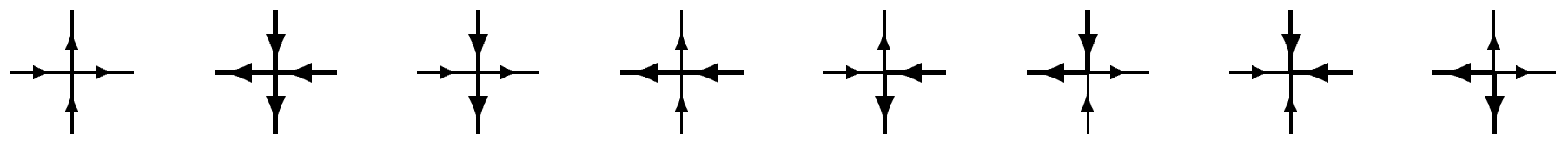

1 2 3 4 5 6 7 8 (4)

Equivalencing the down and left pointing arrows with occupied links (thick arrows in (4)) this model is a self-avoiding random loop model (SALM) with monomer activity $a_1$ and bending rigidity $\eta = a_5 \ldots a_8 = 1/2$. Note that the self-avoiding property follows from the zero weight of vertex two. In the language of the 8-vertex model the system is polarized for all $a_1 > 0$ due to the asymmetry between $a_1$ and $a_2$ and is therefore exposed to an external field. Properly speaking one deals with a 7-vertex model. This fact leads to some computational problems which will be discussed later in more detail.

At these specific couplings the SALM can be solved analytically only in the limit $a_1 = 0$ where it is a critical 6-vertex model [3, 4]. So, the Schwinger model at $\beta = 0$ has at least one critical point at $M = 0$ which is a conformal field theory with central charge $c = 1$ [5]. As already discussed elsewhere [1, 6, 7, 8, 9, 10] there may be a further critical point. If it exists, it may be connected (through a line of critical points) with the second order critical point of the free lattice fermion model at $\kappa = 1/4, \beta = \infty$. In two recent papers this model has been studied, both, at non-zero $\beta$ with Monte Carlo methods [8] and at $\beta = 0$ with computer assisted analytic methods [8, 9] however for very limited lattice sizes. (A table of the available series exceeding the published ones can be found in [11].) In particular the partition function zeros were determined. For the strong coupling Schwinger model (the SALM), from the finite size scaling behaviour of the closest zeros a critical point around $\kappa \approx 0.38$ was conjectured. However, the limited lattice volumes have prevented reliable statements about the order of the transition and its critical exponents, so far. In the approximate analytical approach of [10] arguments for a critical point in the universality class of the Ising model and with conformal charge $c = \frac{1}{2}$were presented and it was pointed out that there exist earlier Monte Carlo results [12] consistent with such a critical point.

To obtain better information on the scaling behaviour one may directly simulate the 7-vertex model by means of Monte Carlo. This has been done in [6] for periodic boundary conditions, i.e. a torus ($\mathbb{Z}_n \times \mathbb{Z}_n$) topology. The results were unsatisfactory for two reasons. First, with the computer resources available at that time, the acquired statistics appears now to be insufficient to obtain the correct scaling behaviour for larger lattices. Furthermore the use of the Ferrenberg-Swendsen multihistogram method may also improve the quality of the results. The second problem is strongly related to the torus topology with its non-trivial homotopy group $\mathbb{Z} \oplus \mathbb{Z}$. This problem was already noticed in [6] but could not be avoided then. In the following section we suggest a different approach.

# 2 Boundary Conditions and Topology

The actual simulation will be discussed later. Let us, however, point out a problem arising in local Monte Carlo updating algorithms on a torus $\mathbb{Z}_n \times \mathbb{Z}_m$. In the Monte Carlo simulation of the model paths are created, evolved and destroyed by *local* updates. Paths which wrap around the torus may be annihilated by joining two of them together, converting such a pair into one path that is closed without running around the torus. It may shrink until it eventually reaches length 4 and then may disappear in a Monte Carlo move.

However, one observes that such a local updating algorithm cannot generate or annihilate one single loop which winds once around the torus in a certain direction. So, given a configuration with either an even or an odd number of loops winding around the torus, this property is conserved during the simulation. On a torus one has therefore four classes of configurations which may be denoted by (0,0), (0,1), (1,0), (1,1). The true equilibrium state is a superposition of configurations of these classes with (at least to us) unknown relative probability weights. In a direct calculation of the partition function (like e.g. the evaluation of the fermionic determinant) all paths are implicitly included with their correct weights.

In [6] the observables had been calculated for these classes separately. Although in the thermodynamic limit $\Lambda \to \infty$ the quantities measured in the different classes approach each other, as expected, on finite lattices there is a non-negligible systematic error due to the fact that the simulation is obviously not ergodic. A solution to this problem requires either a different updating algorithm, which allows to create or destroy loops globally, or different topology or boundary conditions, whereby the problem can be avoided altogether.

Here we follow the second strategy and choose a lattice topology which is isomorphic[1] to $S_2$ (Similar lattices have been used in D=4 studies [13]). The fundamental homotopy group of a sphere is trivial, containing only the unit element. Any loop may be contracted to a point on such a manifold.

Our 2D lattice may be considered as constructed either by gluing together two lattices at their edges or, equivalently, taking the surface of a 3D cubic lattice of size $L \times L \times 1$. Since we want to simulate a vertex model, each site should have four links to nearest neighbours. Thus we introduce additional links connecting each of the four pairs of corner sites. A representative of this pillow– or cushion–like lattice is shown in fig. 1. The advantage lies in the behaviour of closed loops in a local updating process. On this lattice any closed loop may slip over the edges and corners, eventually shrink to a point and disappear. (In fact that updating process defines the homotopy class, which is trivial in this case).

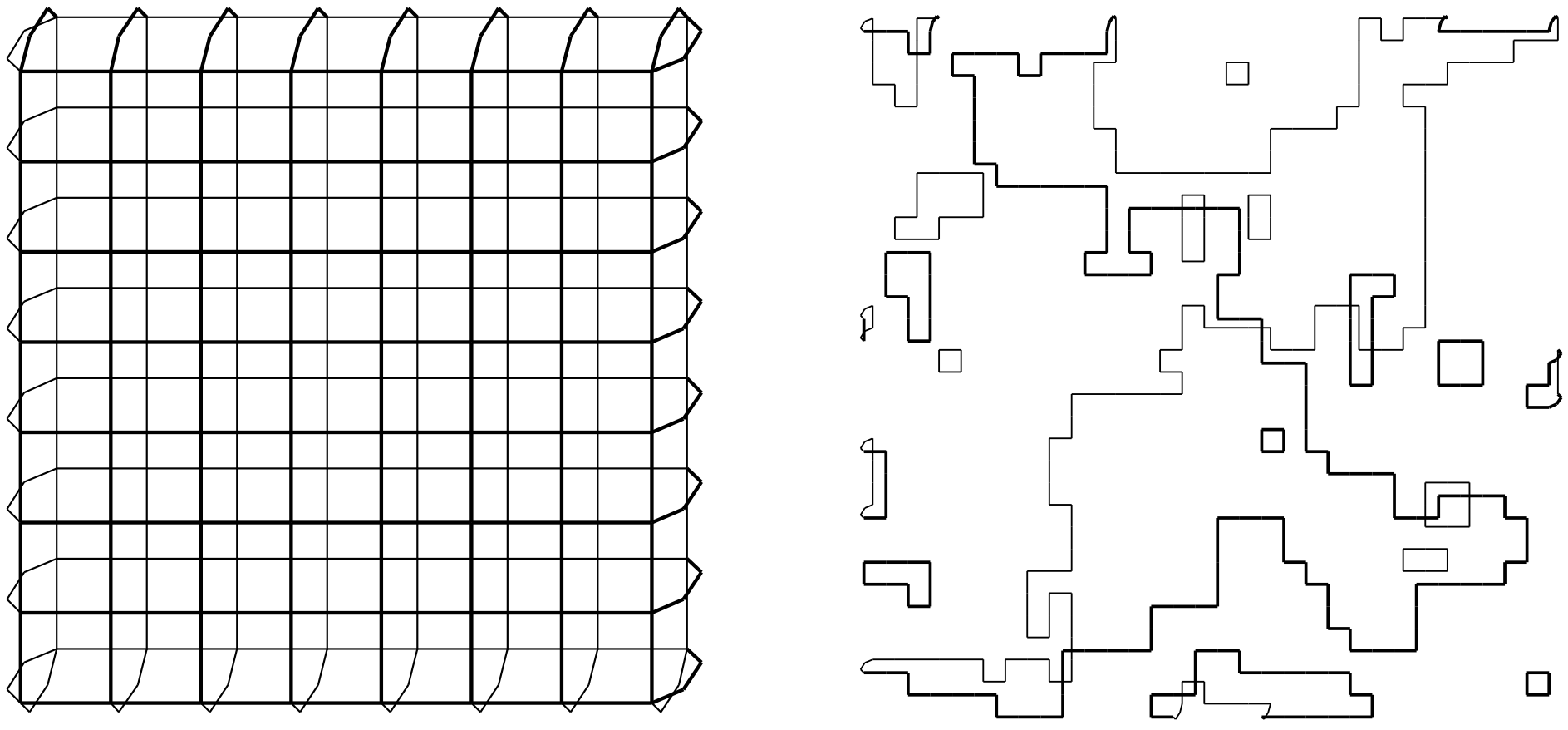

Figure 1: (a) Plot of the lattice topology for the $8 \times 16$ lattice, thick or thin lines are links in the front or in the back. (b) Configuration of a $32 \times 64$ lattice at criticality.

[1]Strictly speaking the terminology used applies only to continuous manifolds; we use it in the sense of embedding the lattices.

# 3 Monte Carlo Simulation and Results

## 3.1 Simulation

We use a Metropolis type [14] updating algorithm and code the configurations in the vertex representation. In the language of (4) a local change of a loop corresponds to a change of all 4 vertices belonging to a fundamental plaquette by flipping the direction of all arrows lying on the bonds of this plaquette. This is a legal mapping from one 8-vertex configuration to another 8-vertex configuration [3]. Such flips may also generate vertices of type 2, allowed in a general 8-vertex model, but excluded in the very constrained 7-vertex model and therefore accepted with *zero* probability. This updating method is as described in [6], with the essential difference being the new lattice topology. The vertices at the corners are treated like the others, except that two of their links (the ones connected to the other corner) are considered duplicated in order to make up plaquettes (where the opposite links are equal).

We simulate the model on lattices $8 \times 16$ up to $128 \times 256$ and determine for each configuration the value $n_1$, which plays a role like the total energy of the system. In fact, summing in (2) over contributions from all other vertex types, the partition function may be written as

$$Z_\Lambda = \sum_{n_1 \in \mathbb{Z}_{|\Lambda|}} \rho_\Lambda(n_1) a_1^{n_1} = \sum_{n_1 \in \mathbb{Z}_{|\Lambda|}} \rho_\Lambda(n_1) M^{2n_1} \ , \tag{5}$$

where $\rho_\Lambda(n_1)$ is the discrete measure of each value of $n_1 \in \mathbb{Z}_{|\Lambda|}$, trivially related to the coefficients of the analytic hopping expansion [8], but evaluated by the Monte Carlo calculation here. By $|\Lambda|$ we denote the lattice (site) volume, in our case $2L^2$.

The individual simulations at various values of $M$ (or $a_1$ or $\kappa$, equivalently) lead to histogram distributions

$$h_\Lambda(n_1, M_i) \propto \rho_\Lambda(n_1) M_i^{2n_1} \tag{6}$$

of the observed values of $n_1$, each of them providing an estimator for the density $\rho_\Lambda(n_1)$. These data may be combined following the multihistogram approach of Ferrenberg and Swendsen [15] producing an optimal estimate for the distribution density. Since this approach is standard by now, we refrain from a further discussion.

## 3.2 Cumulants

From (5) one may find all moments through suitable derivatives

$$\langle n_1^k \rangle = Z_\Lambda^{-1} \frac{\partial^k Z_\Lambda}{\partial(\ln a_1)^k} \; . \tag{7}$$

Employing a notation like for a spin model one may thereby study

$$
\begin{aligned}
&\langle n_1 \rangle && \text{internal energy} && (8)\\
C_{V,\Lambda} = \; & (\langle n_1^2 \rangle - \langle n_1 \rangle^2)/|\Lambda| && \text{specific heat} && (9)\\
V_{BCL} = \; & -\frac{1}{3}\frac{\langle (n_1^2 - \langle n_1^2 \rangle)^2 \rangle}{\langle n_1^2 \rangle^2} && \text{BCL-cumulant [16]} && (10)\\
U_4 = \; & \frac{\langle (n_1 - \langle n_1 \rangle)^4 \rangle}{\langle (n_1 - \langle n_1 \rangle)^2 \rangle^2} && \text{4th order cumulant} && (11)
\end{aligned}
$$

for continuous values of the coupling. At a phase transition one expects typical (peak or minimum) behaviour of the cumulants (9) - (11), depending on the size of the system and on the type of the transition.

Obviously the internal energy (8) is no order parameter, but does serve as an indicator of a possible phase transition. In the language suitable for the underlying Schwinger model, the first two of the above quantities are the chiral density

$$\frac{|\Lambda|}{2} \langle \bar{\psi}_W \psi_W(x) \rangle \tag{12}$$

and the chiral susceptibility. For consistency we will rely here on the spin model terminology.

We run simulations at typically 20-25 values of the coupling $a_1$ with an accumulated statistics of $3 \times 10^6$ configurations for the $8 \times 16$ lattices up to $12 \times 10^6$ configurations for the largest lattice $128 \times 256$. Individual runs at values near the phase transition contributed between $0.12 \times 10^6$ configurations for the small up to $1.2 \times 10^6$ configuration for the largest lattice. The integrated autocorrelation length around the peak position of $C_V$ ranges from $\tau_{int} \simeq 18$ for the smallest lattice up to $\tau_{int} \simeq 1143$ for the largest one. Fig. 2 shows the multihistogram analyses results for $C_V$; we observe a clear peak structure with height increasing with the lattice volume. The value of $C_V/|\Lambda|$ extrapolates to 0 for $L \to \infty$. The other cumulants also indicate clearly that there is a phase transition of 2nd order, in particular $V_{BCL} \to 0$ for $L \to \infty$ (for a 1st order transition it should approach a constant related to the energy gap, see e.g. [17]).

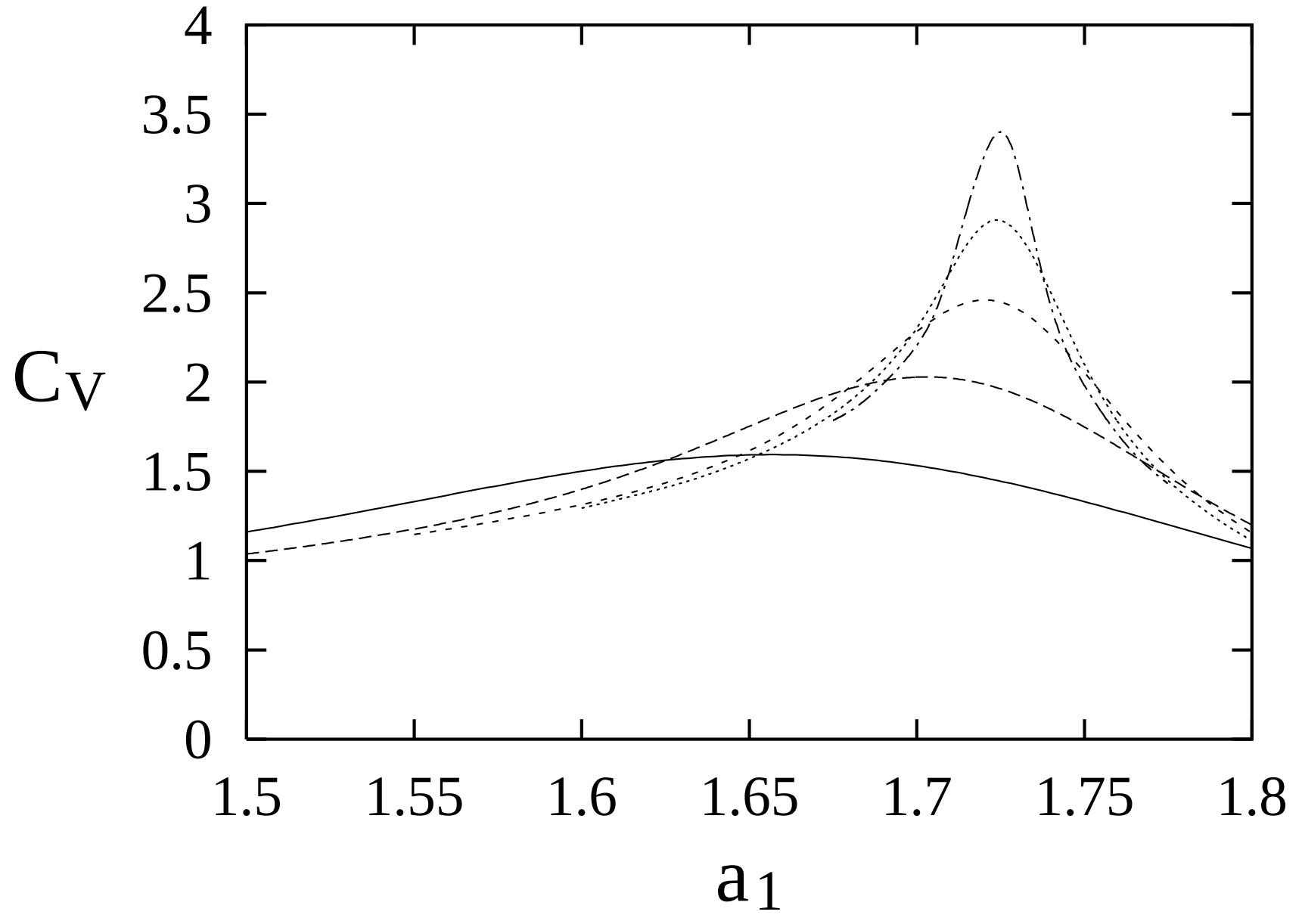

Figure 2: The results for $C_V$ from the multihistogram analysis, for various lattice sizes $L \times 2L$ ($L = 8$, 16, 32, 64, 128)

This first evidence already indicates a phase transition of second or higher order. To simplify the notation, let us introduce the reduced temperature $t \propto (a_1 - a_{1,c}) \propto (M - M_c) \propto (\kappa - \kappa_c)$, vanishing at the critical point. At a second order thermal phase transitions one expects the finite size scaling behaviour

$$C_{V,\Lambda}^{max} \propto L^{\frac{\alpha}{\nu}} , \tag{13}$$

$$|t_L| \propto L^{-\frac{1}{\nu}} . \tag{14}$$

For the critical exponents one has the Josephson relation $\alpha = 2 - D\nu$. To get further information on the finite size scaling we plot the height of the peak in $C_V$ versus $\ln L$ (fig. 3). The functional dependence of the peak hight clearly indicates a logarithmic dependence on $L$. This suggests a critical exponent $\alpha \approx 0$ which, according to Josephson's law, corresponds to $\nu = 1$.

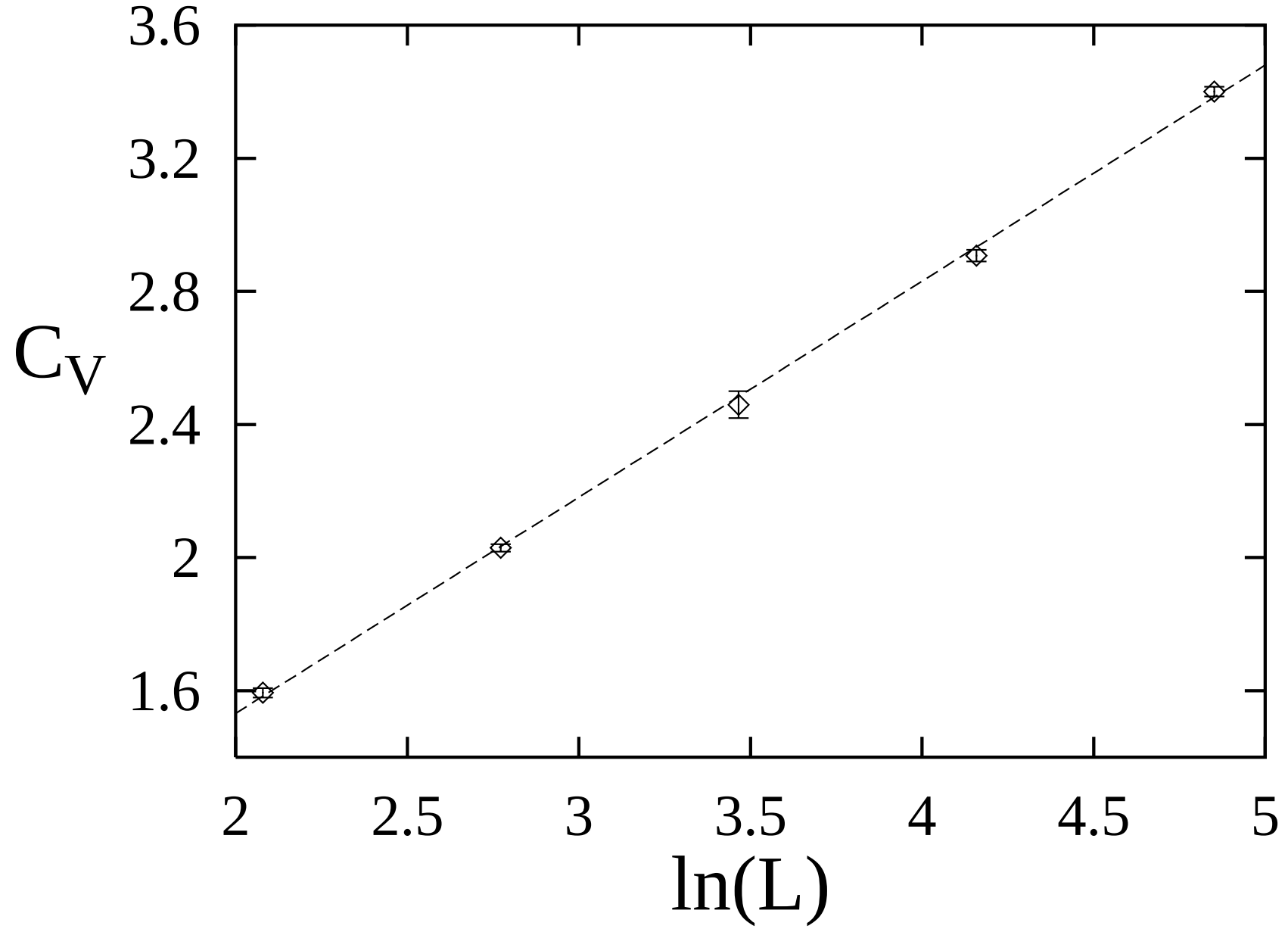

Figure 3: The peak values of $C_V$ vs. $\ln L$.

We define pseudocritical couplings via the peak position of $C_V$ and the different minima positions of $V_{BCL}$ and $U_4$. In the infinite volume limit all these pseudocritical couplings should converge to a common critical value $a_{1,c}$. A finite size scaling ansatz according

$$a_1(L) = a_{1,c} + \frac{c}{L^\epsilon} \ , \tag{15}$$

with a common value of $a_1$ and $\epsilon$ may be fitted to all three cumulants. Although this fit is of good quality ($\chi^2/d.f. \simeq 2.2$) the result for $\nu = 1/\epsilon \simeq 0.58$ is close to the value expected for a first order transition $\nu = 1/D$ and is in contradistinction to the discussed evidence for the peak value of $C_V$. In the next section this discrepancy will be resolved. Anticipating the result and in agreement with our observation for the peak value of $C_V$ we therefore prefer another parameterization.

It turns out that the finite size behaviour may be parametrized consistently by

$$a_1(L) = a_{1,c} + \frac{c}{L} + \frac{d}{L^2} \ . \tag{16}$$

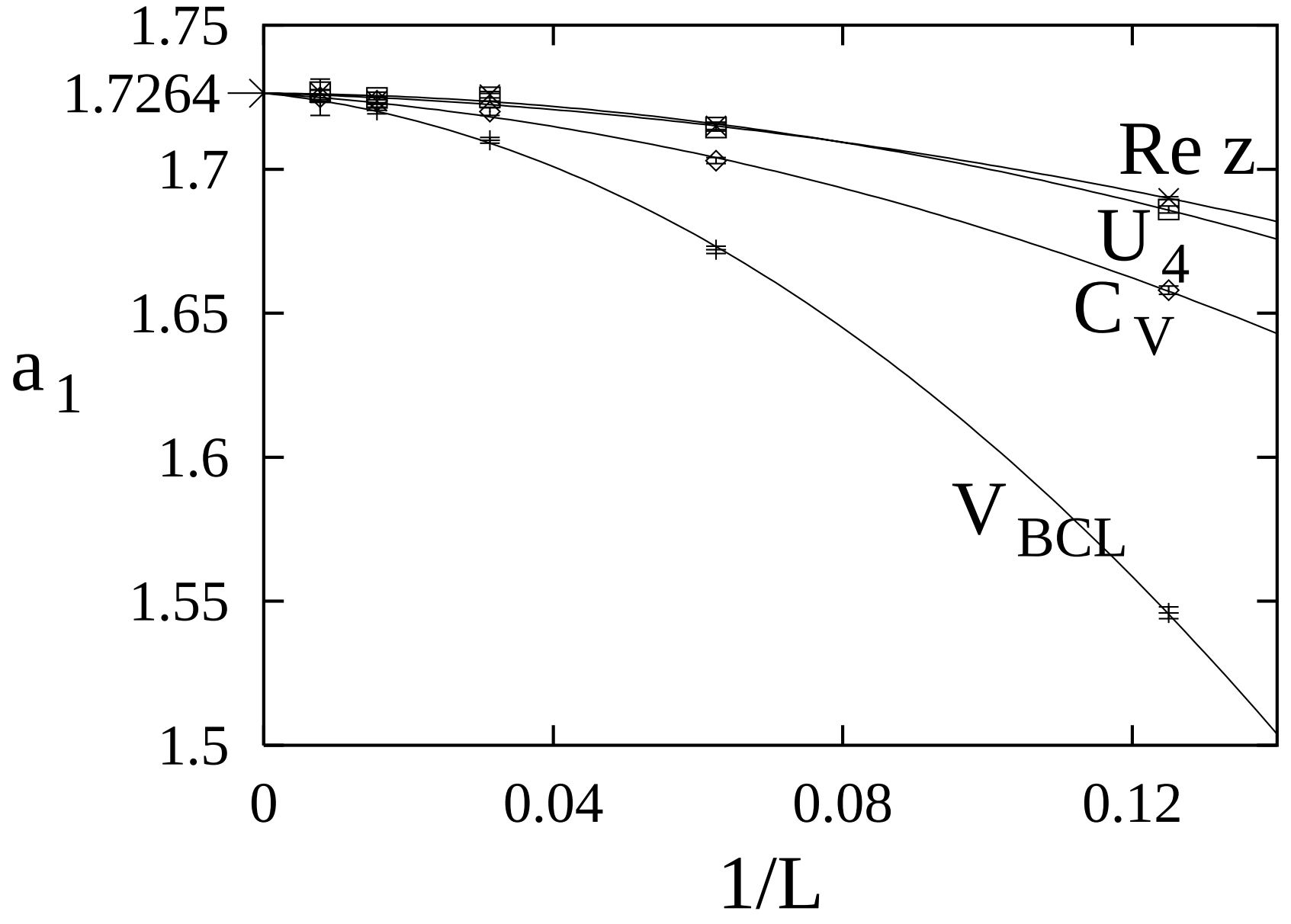

Figure 4: The pseudocritical values for $C_V$, $V_{BCL}$ and $U_4$ as well as the real pat of the position of the closest partition function zero Re$(z)$ fitted according (16); it is obvious, that the term linear in $1/L$ is very small.

A joint fit to these pseudocritical data (and to the real part of the position of the closest complex zero of the partition function, to be discussed below), allowing for different coefficients $c$ and $d$ but identical $a_{1,c}$ gives a value $a_{1,c} = 1.7264(5)$ or $\kappa_c = 0.3805(1)$, correspondingly ($\chi^2/d.f. \simeq 2.3$). The constant $c$ comes out *very small* as compared to $d$ (cf. fig. 4).

This behaviour resembles that of similar quantities in other studies of 4D [13] and 2D models on spherelike lattices. In fact, the term $O(\frac{1}{L^2})$ is readily explained from the "boundary" terms in the free energy. In our lattice topology the curvature is concentrated in the 8 corners, a source of non-homogeneity, albeit suppressed in the free energy per unit volume like a $1/|\Lambda| \propto 1/L^2$ term. The real surprise is the smallness of the leading finite size scaling term, which forbids a reliable determination of the critical exponent $\nu$ from these observables.

| L | Re $\kappa_0$ | Im $\kappa_0$ |
|---|---|---|
| 8 | 0.38463(6) | 0.02943(7) |
| 16 | 0.38180(14) | 0.01456(9) |
| 32 | 0.38060(12) | 0.00729(33) |
| 64 | 0.38085(5) | 0.00378(24) |
| 128 | 0.38052(9) | 0.00176(6) |

Table 1: The positions of the closest zeros in the complex $\kappa$ variable.

### 3.3 Partition function zeros

This apparent puzzle finds its solution in the closer inspection of the behaviour of the partition function zeros. Eq. 5 defines the partition function also at complex values of $a_1$ (or $M$ or $\kappa$). Due to the multihistogram method the density is determined with sufficient quality to reliably determine the positions of the zeros closest to the real axis of the coupling parameter (cf. a similar study in [18]). As observed by Yang and Lee [19] for the odd field coupling and further elaborated by Fisher [20] for the temperature coupling for lattices of finite volume these zeros have non-vanishing imaginary parts.

For better comparison with earlier work [8, 9] we analyze the position of the zero in the variable $\kappa$. Table 1 and fig. 5 give – for the lattices studied – those zeros in the complex $\kappa$ variable closest to the real axis.

The finite size scaling relation (13) implies for the zero closest to the real axis

$$\mathrm{Im}\ \kappa_0 \simeq L^{-\frac{1}{\nu}} \tag{17}$$

and gives another observable for the critical exponent $\nu$. Fig. 6 exhibits excellent finite size scaling; a fit according (17) gives a value of $\nu = 0.986(7)$ (with a $\chi^2/d.f. \simeq 0.3$). This is in perfect agreement with $\alpha = 0$, i.e. with the result for the logarithmic behaviour of $C_V$.

Whereas the imaginary part scales nicely, one sees (fig. 5) that the real part changes very little and therefore has little chance to exhibit the leading scaling behaviour. We think that this explains the earlier mentioned scaling behaviour of the pseudocritical values derived from the cumulants. All peak- or minima-positions are related essentially to the real part of the nearby partition function zeros. This also demonstrates the inherent danger of relying on some observables exclusively. Only studies of several observables eventu-ally produce a coherent picture. Also, the imaginary part of the partition function zero proves to be of particular value in the analysis.

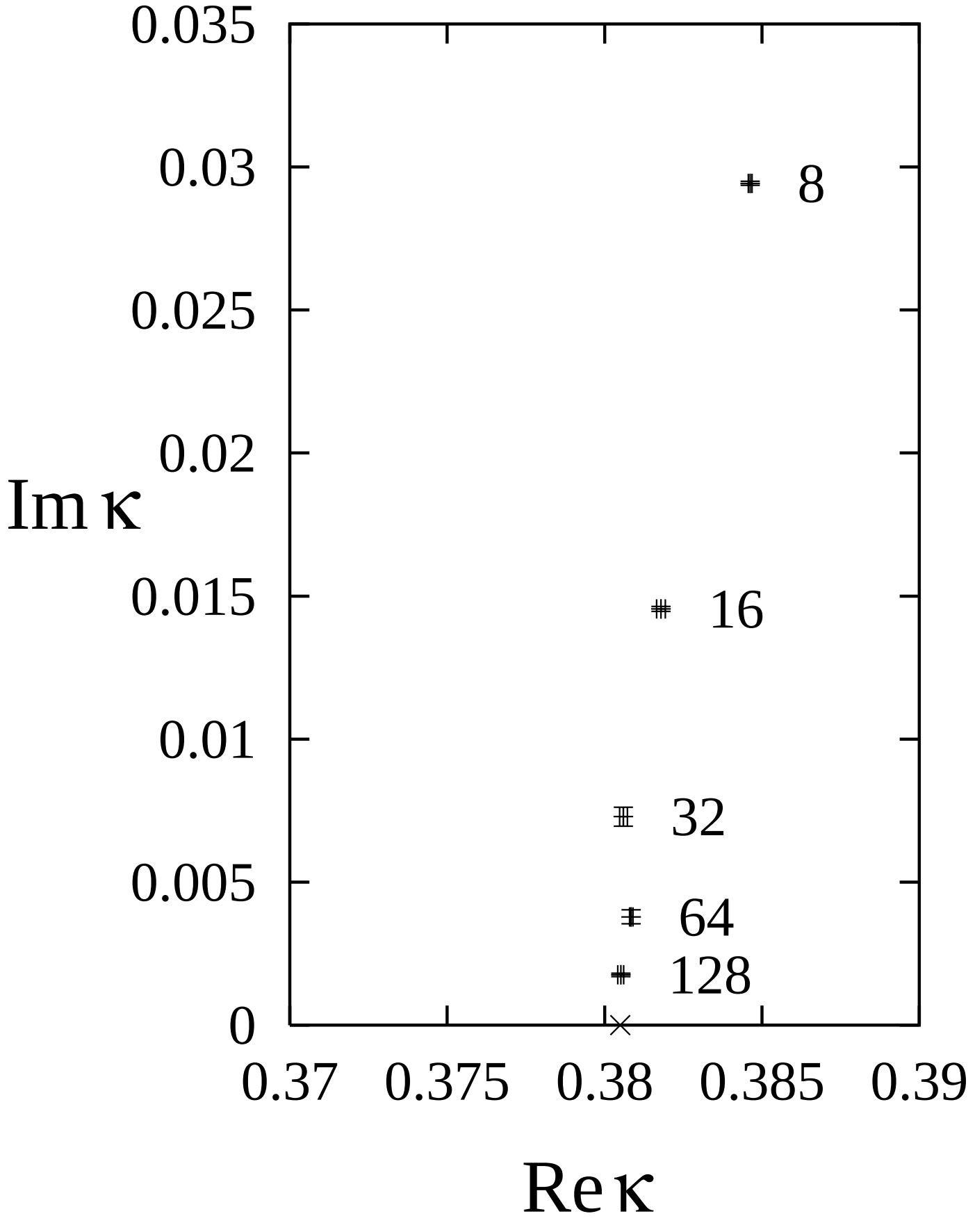

Figure 5: The closest zeros of the partition function in the complex $\kappa$-plane for various lattice sizes.

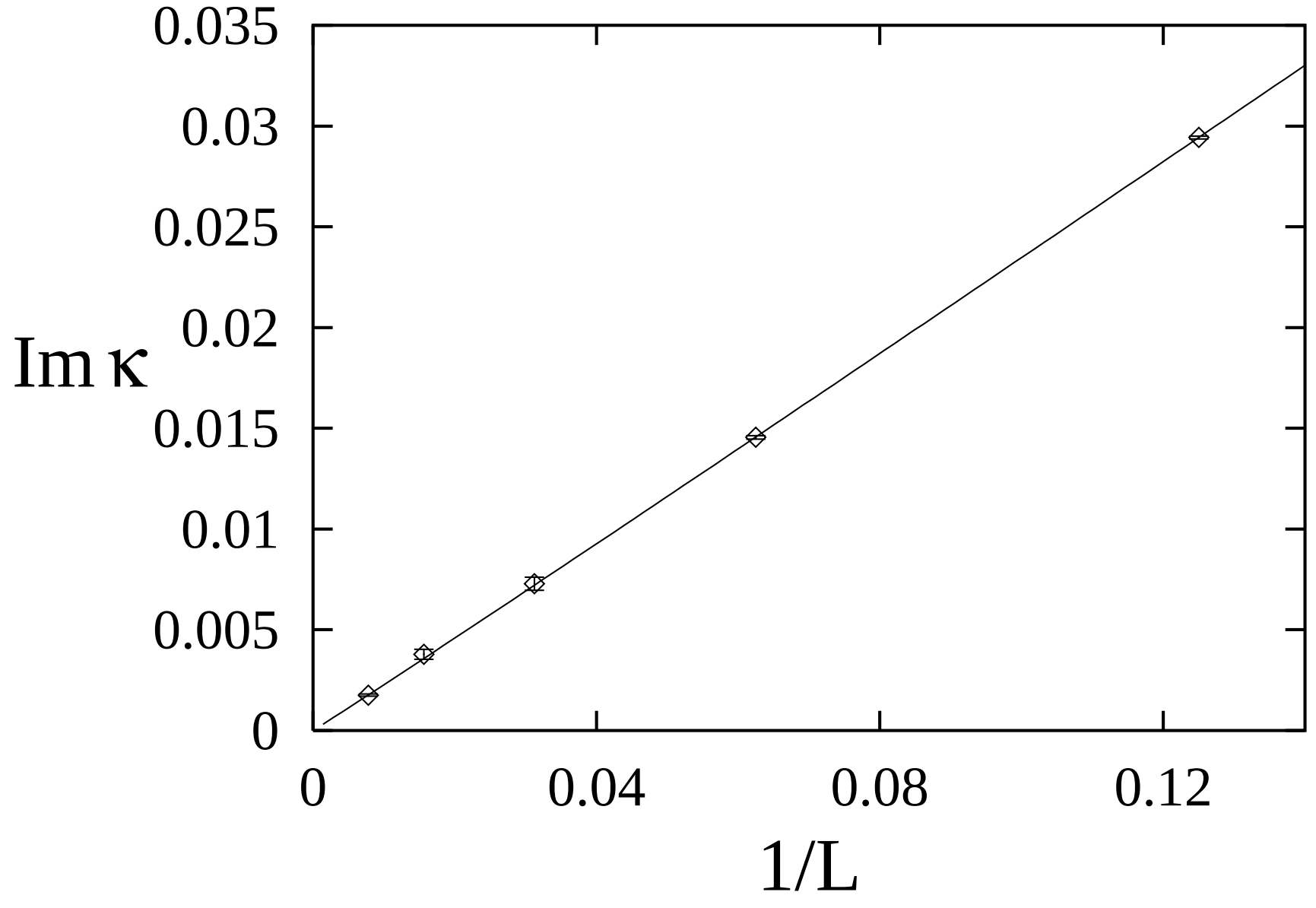

Figure 6: The imaginary part of the partition function zeros closest to the real axis in the $\kappa$ variable.

All this evidence suggests that the one flavor lattice regularized strongly coupled Schwinger Model with Wilson fermions has, in addition to a transition at $\kappa_c = \infty$, a second order critical point at $\kappa = 0.3805(1)$ which is in the universality class of the 2D Ising model. Defining a continuum field theory at this transition gives a model of free fermions (or, equivalently, the Ising model: a model of strongly interacting bosons). In the equivalent vertex model picture we find a 2nd order phase transition at $a_1 = 1.7264(5)$ with a critical exponent $\nu \simeq 1$.

# 4 Conclusions

Let us conclude with some remarks on the relevance of our results for the lattice Schwinger model (n.b. with *one* fermion flavor). The presented study

demonstrates – quite convincingly, we think – that the model has a 2nd order phase transition at $\beta = 0$, $\kappa = 0.3805(1)$ with critical exponent $\nu \simeq 1$. This is in agreement with a recent approximate analytical study of the SALM [10]. The model also has a (free fermion) phase transition at $\beta = \infty$, $\kappa = 0.25$ with $\nu = 1$. Both transitions are in the universality class of the Ising– or the free-fermion model. Other studies [7, 8, 9] have given some evidence, that there may be in fact a critical line in $\kappa$ for all $\beta \in [0, \infty)$ (supported also by a recent study of the non-compact version [21]).

There are two scales in the model, one scale $\xi_\kappa \propto 1/(\kappa - \kappa_c(\beta))$ is governed by $\kappa$ and corresponds to $\nu_\kappa = 1$ (free fermions) and the other scale is governed by the gauge field $\beta$ with $\xi_\beta \propto \sqrt{\beta}$. The limit to the continuum massive (and massless) Schwinger model then will be found in the approach $\beta \to \infty$, $\kappa \to \kappa_c(\beta)$ along curves

$$\sqrt{\beta}\,(\kappa - \kappa_c(\beta)) = const.\ , \tag{18}$$

where the constant is related to the mass of the boson in the continuum theory. The massless continuum Schwinger model will be approached along the critical curve $\kappa = \kappa_c(\beta)$. A different approach will not reconstruct the continuum Schwinger model but the critical Ising model or equivalently free fermions.

**Acknowledgement:**

We are indebted to M. Salmhofer for many discussions. We also want to gratefully acknowledge various interesting discussions and e-mail conversations with V. Azcoiti, Ch. Gattringer, H. Grosse, H. Scharnhorst and E. Seiler.